\documentclass[a4paper,11pt]{article}

\usepackage[utf8]{inputenc}
\usepackage[T1]{fontenc}
\usepackage{amsfonts,amsmath,algorithm,algpseudocode,multicol}
\usepackage[colorinlistoftodos]{todonotes}
\newcommand{\RR}{\mathbb{R}}
\newcommand{\QQ}{\mathbb{Q}}

\newcommand{\none}{\textsf{none}}
\newcommand{\true}{\textsf{true}}
\newcommand{\false}{\textsf{false}}
\newcommand{\ignore}[1]{{}}

\newcommand{\Registered}{\textsuperscript{\textregistered}}
\graphicspath{{figures/}{xp/}}
\usepackage{paralist,afterpage}
\usepackage{hyperref}
\usepackage{subfig}
\usepackage{amsthm}

\theoremstyle{remark}
\newtheorem{example}{Example}

\newenvironment{contexample}{\addtocounter{example}{-1}\begin{example}[continued]}{\end{example}}

\title{Parallel parametric linear programming solving, and application to polyhedral computations}

\author{Camille Coti \and David Monniaux \and Hang Yu}

\begin{document}

\maketitle

\begin{abstract}
Parametric linear programming is central in polyhedral computations and in certain control applications.
We propose a task-based scheme for parallelizing it, with quasi-linear speedup over large problems.
\end{abstract}

\section{Introduction}
\label{sec:intro}
A \emph{convex polyhedron}, or \emph{polyhedron} for short here, in dimension $n$ is the solution set over $\QQ^n$ (or, equivalently,
$\RR^n$) of a system of inequalities (with integer or rational coefficients).
Polyhedra in higher dimension are typically used to enclose the reachable states of systems whose state can be expressed, at least partially, as a vector of reals or rationals; e.g. hybrid systems or software~\cite{DBLP:conf/popl/CousotH78}.

The conventional approaches for polyhedral computations are the \emph{dual description} (using both vertices and faces) and \emph{Fourier-Motzkin elimination}.
They both suffer from high complexity on relevant cases.
We instead express image, projection, convex hull etc. as solutions to \emph{parametric linear programmings}, where parameters occur linearly within the objective.
A solution to such a program is a quasi-partition of the space of parameters into polyhedra, with one optimum associated to each polyhedron.
The issue is how to compute this solution efficiently.
In this article, we describe how we parallelized our algorithm.




\section{Sequential algorithms}
\label{sec:algo}
Here we are leaving out how polyhedral computations such as projection and convex hull can be reduced to parametric linear programming --- this is covered in the literature \cite{Jones+JOTA08-On-polyhedral-projections-and-parametric-programming,Marechal_Monniaux_Perin_SAS2017} --- and focus on solving the parametric linear programs.

\subsection{Non-parametric linear programming (LP)}
A linear program with $n$ unknowns is defined by a system of equations $AX = B$, where $A$ is an $m \times n$ matrix; a solution is a vector $X$ such that $X \geq 0$ on all coordinates and $AX = B$. %
The program is said to be \emph{feasible} if it has at least one solution, \emph{infeasible} otherwise.
In a non-parametric linear program one considers an objective $C$: one wants the solution that maximizes $C^T X$.
The program is deemed \emph{unbounded} if it is feasible yet it has no such optimal solution.

\begin{example}
\label{ex:polygon}
  Consider the polygon $P$ defined by $x_1 \geq 0$, $x_2 \geq 0$, $3x_1-x_2 \leq 6$, $-x_1+3x_2 \leq 6$.
  Define $x_3 = 6 - 3x_1 + x_2$ and $x_4 = 6 + x_1 - 3x_2$.
  Let $X = (x_1,x_2,x_3,x_4)$, and then $P$ is the projection onto the first two coordinates of the solution set of $AX=B \land X \geq 0$ where
  $A = \left[\begin{smallmatrix}
      1 & -3 & 0 & -1 \\
      -3 & 1 & -1 & 0 
    \end{smallmatrix}\right]$
    and $B = \left[\begin{smallmatrix} 6 \\ 6 \end{smallmatrix}\right]$.
\end{example}

An LP solver takes as input $(A,B,C)$ and outputs ``infeasible'', ``unbounded'' or an optimal solution.
Most solvers work with floating-point numbers and their final answer may be incorrect: they may answer ``infeasible'' whereas the problem is feasible, or give ``optimal solutions'' that are not solutions, or not optimal.

In addition to a solution $X^*$, solvers also provide the associated \emph{basis}:
$X^*$ is defined by setting $n-m$ of its coordinates to $0$ (known as \emph{nonbasic variables}) and solving for the other coordinates (known as \emph{basic variables}) using $AX^* =B$,
and the solver provides the partition into basic and nonbasic variables it used.
If a floating-point solver is used, it is possible to reconstruct an exact rational point $X^*$ using that information and a library for solving linear systems in rational arithmetic.
One then checks whether it is truly a solution by checking $X^* \geq 0$.

The optimal basis also contains a proof of optimality of the solution.
We compute the objective function $C^T X$ as $\sum_{i \in N} \alpha_i X_i+c$ where $N$ is the set of indices of the nonbasic variables and $c$ is a constant,
and conclude that the solution obtained by setting these nonbasic variables to $0$ is maximal because all the $\alpha_i$ are nonpositive.
If $X^*$ is not a solution of the problem ($X^* \geq 0$ fails) or is not optimal, then we fall back to an exact implementation of the simplex algorithm.

\begin{contexample}
  Assume the objective is $C = \begin{bmatrix} 1 & 1 & 0 & 0\end{bmatrix}$, that is, $C^T X = x_1 + x_2$.
  From $AX = B$ we deduce
  $x_1 = 3 -\tfrac{3}{8} x_3 - \tfrac{1}{8} x_4$ and
  $x_2 = 3 -\tfrac{1}{8} x_3 - \tfrac{3}{8} x_4$.
  Thus $x_1 + x_2 = 6 - \frac{1}{2} x_3 - \frac{1}{2} x_4$.

  Assume $x_3$ and $x_4$ are nonbasic variables and thus set to $0$, then
  $X^* = (x_1,x_2,x_3,x_4) = (3, 3, 0, 0)$.
  It is impossible to improve upon this solution: as $X \geq 0$, changing the values of $x_3$ and $x_4$ can only decrease the objective $o = 6 - \frac{1}{2} x_3 - \frac{1}{2} x_4$.
  This expression of $o$ from the nonbasic variables can be obtained by linear algebra once the partition into basic and nonbasic variables is known.
\end{contexample}


While the optimal value $C^T X^*$, if it exists, is unique for a given $(A,B,C)$, there may exist several $X^*$ for it, a situation known as \emph{dual degeneracy}.
%
%
The same $X^*$ may be described by different bases, a situation known as \emph{primal degeneracy}, happening when more than $n-m$ coordinates of $X^*$ are zero, and thus some basic variables could be used as nonbasic and the converse.

\subsection{Parametric linear programming (PLP)}
For a \emph{parametric} linear program, we replace the constant vector $C$ by $C_0  + \sum_{i=1}^k \mu_i C_i$ where the $\mu_i$ are parameters.%
\footnote{There exists another flavor of PLP with parameters in the right-hand sides of the constraints.}
When the $\mu_i$ change, the optimum $X^*$ changes.
Assume temporarily that there is no degeneracy. Then, for given values of the $\mu_i$, the problem is either unbounded, or there is one single optimal solution~$X^*$.
It can be shown that the region of the $(\mu_1,\dots,\mu_k)$ associated to a given optimum $X^*$ is a convex polyhedron (if $C_0=0$, a convex polyhedral cone), and that these regions form a quasi partition of the space of parameters (two reegions may overlap at their boundary, but not in their interior) \cite{Jones+JOTA08-On-polyhedral-projections-and-parametric-programming,Jones+Automatica2007-Multiparametric-linear-programming-with-applications-to-control,Marechal_Monniaux_Perin_SAS2017}.
The output of the parametric linear programming solver is this quasi-partition, and the associated optima---in our applications, the problem is always bounded in the optimization directions, so we do not deal with the unbounded case.

Let us see in more details about how to compute these regions.
We wish to attach to each basis (at least, each basis that is optimal for at least one vector of parameters) the region of parameters for which it is optimal.

\setcounter{example}{0} 
\begin{example}[continued]
Instead of $C=\begin{bmatrix} 1 & 1 & 0 & 0\end{bmatrix}$ we consider $C=\begin{bmatrix} \mu_1 & \mu_2 & 0 & 0 \end{bmatrix}$.
Let us now express $o = C^T X$ as a function of the nonbasic variables $x_3$ and $x_4$:
\begin{equation}
o = (3 \mu_1 + 3 \mu_2) +
    \left(-\tfrac{3}{8} \mu_1 -\tfrac{1}{8} \mu_2\right) x_3 +
    \left(-\tfrac{1}{8} \mu_1 -\tfrac{3}{8} \mu_2\right) x_4
\end{equation}
The coefficients of $x_3$ and $x_4$ are nonpositive if and only if
$3\mu_1 + \mu_2 \geq 0$ and $\mu_1 + 3\mu_2 \geq 0$, which define
the cone of optimality associated to that basis and to the optimum
$X^* = (3,3,0,0)$.
\end{example}

The description of the optimality polyhedron by the constraints obtained from the sign conditions in the objective function may be redundant: containing constraints that can be removed without changing the polyhedron.
Our procedure~\cite{DBLP:conf/vmcai/MarechalP17} for removing redundant constraints from the description of a region $R_1$ also provides a set of vectors outside of $R_1$,
a feature that will be useful.

Assume now we have solved the optimization problem for a vector of parameters $D_1$, and obtained a region $R_1$ in the parameters (of course, $D_1 \in R_1$).
We store the set of vectors outside of $R_1$ provided by the redundancy elimination procedure into a ``working set'' $W$ to be processed, choose $D_2$ in it.
We compute the region $R_2$ associated to $D_2$.
Assume that $R_2$ and $R_1$ are adjacent, meaning that they have a common boundary.
We get vectors outside of $R_2$ and add them to~$W$.
We pick $D_3$ in $W$, check that it is not covered by $R_1$ or $R_2$, and, if it is not, compute $R_3$, etc.
The algorithm terminates when $W$ becomes empty, meaning the $R_1,\dots$ produced form the sought quasi-partition.

This simplistic algorithm can fail to work because it assumes that it is discovering the adjacency relation of the graph.
The problem is that, if we move from a region $R_i$ to a vector $D_j \notin R_i$, it is not certain that the region $R_j$ generated from $D_j$ is adjacent to $R_i$ --- we could miss some intermediate region.
We modify our traversal algorithm as follows.
The working set contains pairs $(R,D')$ where $R$ is a region and $D' \notin R$ a vector (there is a special value \none\ for $R$).
The region $R'$ corresponding to $D'$ is computed.
If $R$ and $R'$ are not adjacent, then a vector $D''$ in between $R$ and $R'$ is computed, and $(R,D'')$ added to the working set.
This ensures that we obtain a quasi-partition in the end.
Additionally, we obtain a spanning tree of the region graph, with edges from $R$ to $R'$.

The last difficulty is degeneracy.
We have so far assumed that each optimization direction corresponds to exactly one basis.
In general this is not the case, and the interiors of the optimality regions may overlap.
This hinders performance.
The final result is no longer a quasi-partition, but instead just a covering of the parameter space---enough for projection, convex hull etc. being correct.


\section{Parallel parametric linear programming}
\begin{algorithm}[t]
  \caption{Concurrent push on the shared region structure.}
  \label{algo:push_region}
  \vspace{-1em}
  
  \begin{multicols}{2}
  \begin{algorithmic}
    \Procedure{push\_region}{$R$}
  \State \textbf{atomic} ($i \gets n_{\mathrm{fill}}$; $n_{\mathrm{fill}} \gets n_{\mathrm{fill}} +1$)
  \State $\mathit{regions}[i] \gets R$
  \While{$n_{\mathrm{ready}} < i$}
    \State possibly use a condition variable instead of spinning
  \EndWhile \Comment{$n_{\mathrm{ready}} = i$}
  \State \textbf{atomic} $n_{\mathrm{ready}} \gets i+1$
  \EndProcedure
  \end{algorithmic}
\end{multicols}\vspace{-1em}

\end{algorithm}

\begin{algorithm}[t]
  \caption{Task for parallel linear programming solver.}

  $\mathit{push\_tasks}$ adds new tasks to be processed (different under TBB and OpenMP).

  $\mathit{test\_and\_insert}(T,x)$ checks whether $x$ already belongs to the hash table $T$, in which case it returns $\true$; otherwise it adds it and returns $\false$. This operation is atomic.\vspace{-1em}
  
  \begin{multicols}{2}%
  \begin{algorithmic}%
  \Procedure{process\_task}{$(R_{\mathrm{from}},D)$}
  \State $R_{\mathrm{cov}} \gets \mathit{is\_covered}(D,\mathit{regions})$
  \If{$R_{\mathrm{cov}} == \none$}
    \State $\mathit{basis} \gets \mathit{float\_lp}(A,B,C(D))$
    \If{$\neg \mathit{test\_and\_insert}(\mathit{bases}, \mathit{basis})$}
      \State $X^* \gets \mathit{exact\_point}(\mathit{basis})$
      \State $o \gets \mathit{exact\_objective}(\mathit{basis})$
      \If{$\neg (X^* \geq 0 \land o \leq 0)$}
        \State $(\mathit{basis},X^*) \gets \mathit{exact\_lp}(A,B,C(D))$
      \EndIf
      \State $S \gets \mathit{sign\_conditions}(\mathit{basis})$
      \State $R \gets eliminate\_redundancy(S)$
      \For{each constraint $i$ in $R$}
        \State $D_{\mathrm{next}} \gets \mathit{compute\_next(R,i)}$
        \State $\mathit{push\_tasks}(D_{\mathrm{next}})$
      \EndFor
      \State $\mathit{push\_region(R, X^*)}$
      \State $R_{\mathrm{cov}} \gets R$
    \EndIf
  \EndIf
  \If{$\neg \mathit{are\_adjacent}(R_{\mathrm{from}},R_{\mathrm{cov}})$}
    \State $D' \gets \mathit{midpoint}(R_{\mathrm{from}},R_{\mathrm{cov}},D)$
    \State $W \gets W \cup \{ (R_{\mathrm{from}}, D') \}$
  \EndIf
  \EndProcedure
\Statex
\Procedure{$\mathit{is\_covered}$}{$D,\mathit{regions})$}
\For{$i \in 0\dots n_{\mathrm{ready}}-1$}
   \Comment{$n_{\mathrm{ready}}$ to be read at every loop iteration}
   \State $(R,X^*) \gets \mathit{regions}[i]$
   \If{$D$ covered by $R$}
     \State \textbf{return}($R$)
   \EndIf
\EndFor
\State \textbf{return}(\none)
\EndProcedure
  \end{algorithmic}
\end{multicols}\vspace{-1em}

\end{algorithm}

\label{sec:para}

Our algorithms are designed in a task-based execution model. 
%
The sequential algorithm executes tasks taken from a working set, which can themselves spawn new tasks.
In addition, it maintains the set $\mathit{regions}$ of regions already seen, used:
\begin{inparaenum}[i)]
\item for checking if a vector $D$ belongs to a region already covered ($\mathit{is\_covered}$);
\item for checking adjacency of regions;
\item for adding new regions found.
\end{inparaenum}
Therefore, in a parallel task model, this algorithm is
straightforwardly parallel. The regions are inserted into a concurrent
array. We investigated two task scheduling strategies. A \emph{static}
approach starts all the available tasks, waits for them to complete
and collects all the new tasks $(R,D)$ into the working set, until no
new task is created and the working set is empty. A \emph{dynamic}
approach allows adding new tasks to the working set dynamically and
runs the tasks until that set is empty.

The number of tasks running to completion (not aborted early due to a test) is the same as the number of generated regions.
%
The $\mathit{is\_covered}(D,\mathit{regions})$ loop can be easily parallelized as well.
We opted against it as it would introduce a difficult-to-tune second level of parallelism.

We implemented these algorithms using Intel's Thread Building Blocks
(TBB \cite{TBB}) and 
OpenMP tasks~\cite{OpenMP_4.5}, both providing a task-based
parallelism model with different features.

The dynamic task queue can be implemented using TBB's
\verb|tbb::parallel_do|, which dynamically schedules tasks from the
working set on a number of threads. The static scheduling approach can
simply be implemented by a task synchronization barrier (such as
OpenMP's barrier).

That first implementation of the dynamic task scheduling approach was slow.
The working set often contained tasks such that the regions generated from them were the same, leading to redundant computations.
The workaround was to add a hash table
storing the set of bases (each being identified by the ordered set of its basic variables) that have been or are currently being processed. 
A task will abort after solving the floating-point linear program if it finds that its basis is already in the table.



\section{Performance evaluation}
\label{sec:xp}

\ignore{
\begin{figure}[tp]
{\centering\includegraphics{omp_P_120_0_50_0_proj1_rennes_be0be063}}
\caption{Computation time and speedup for 50 projections of polyhedra in dimension 120. Each parametric linear program has 3460--3715 regions.}
\label{fig:large_polyhedra}
\end{figure}

\begin{figure}[tp]
{\centering\includegraphics{auto_P_29_15_16_3_be0be063_paranoia-8_rennes_grid5000_fr}}
\caption{Computation time and speedup for smaller, simpler polyhedra.}   
\label{fig:small_polyhedra}
\end{figure}
}

\begin{figure}[tp]
    \subfloat[2 dimensions projected]{
      \includegraphics[width=0.48\textwidth,height=3.8cm]{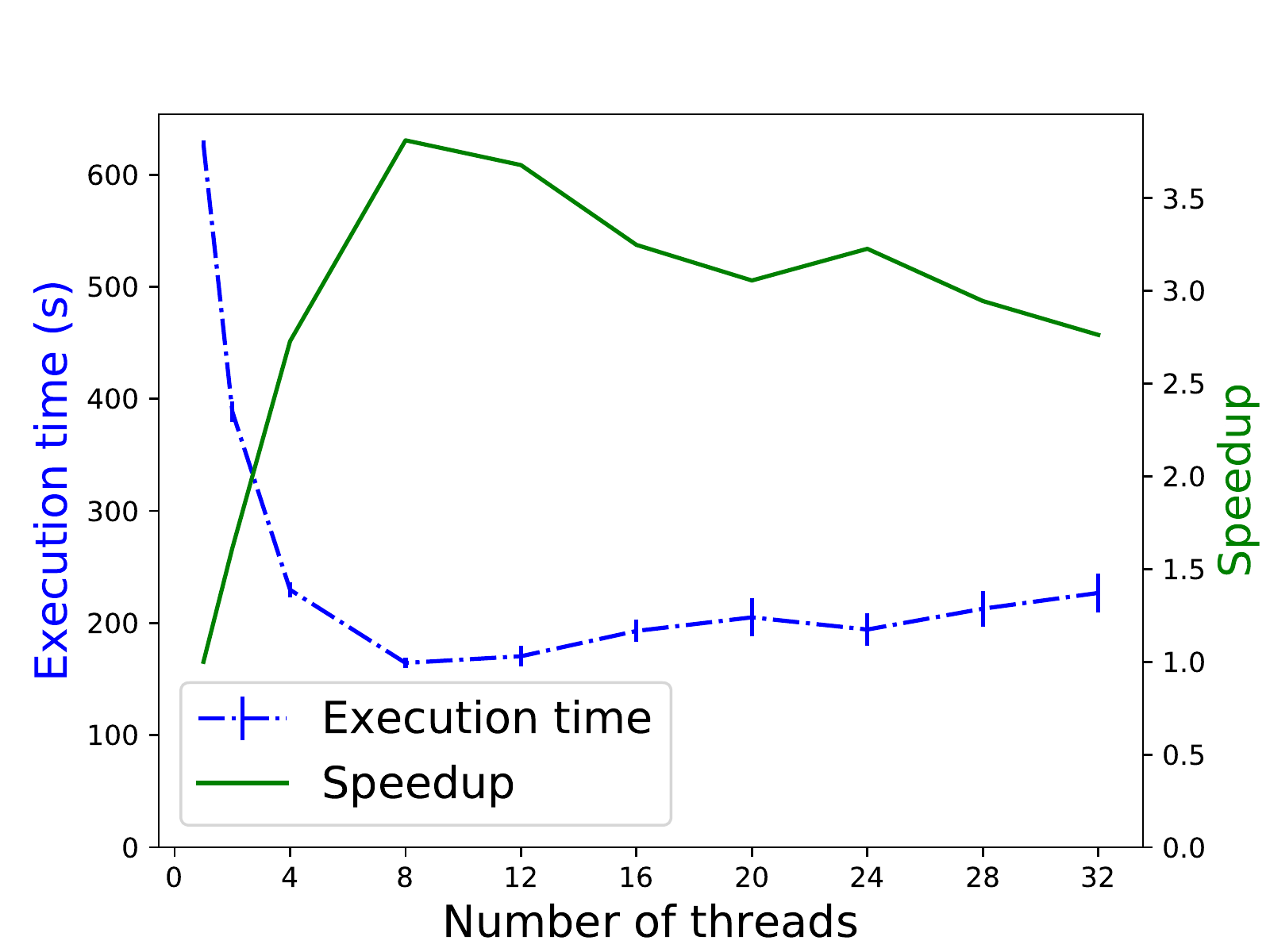}
      \label{fig:xp:9_0_20_16_2}
    }\hfill
    \subfloat[5 dimensions projected]{
      \includegraphics[width=0.48\textwidth,,height=3.8cm]{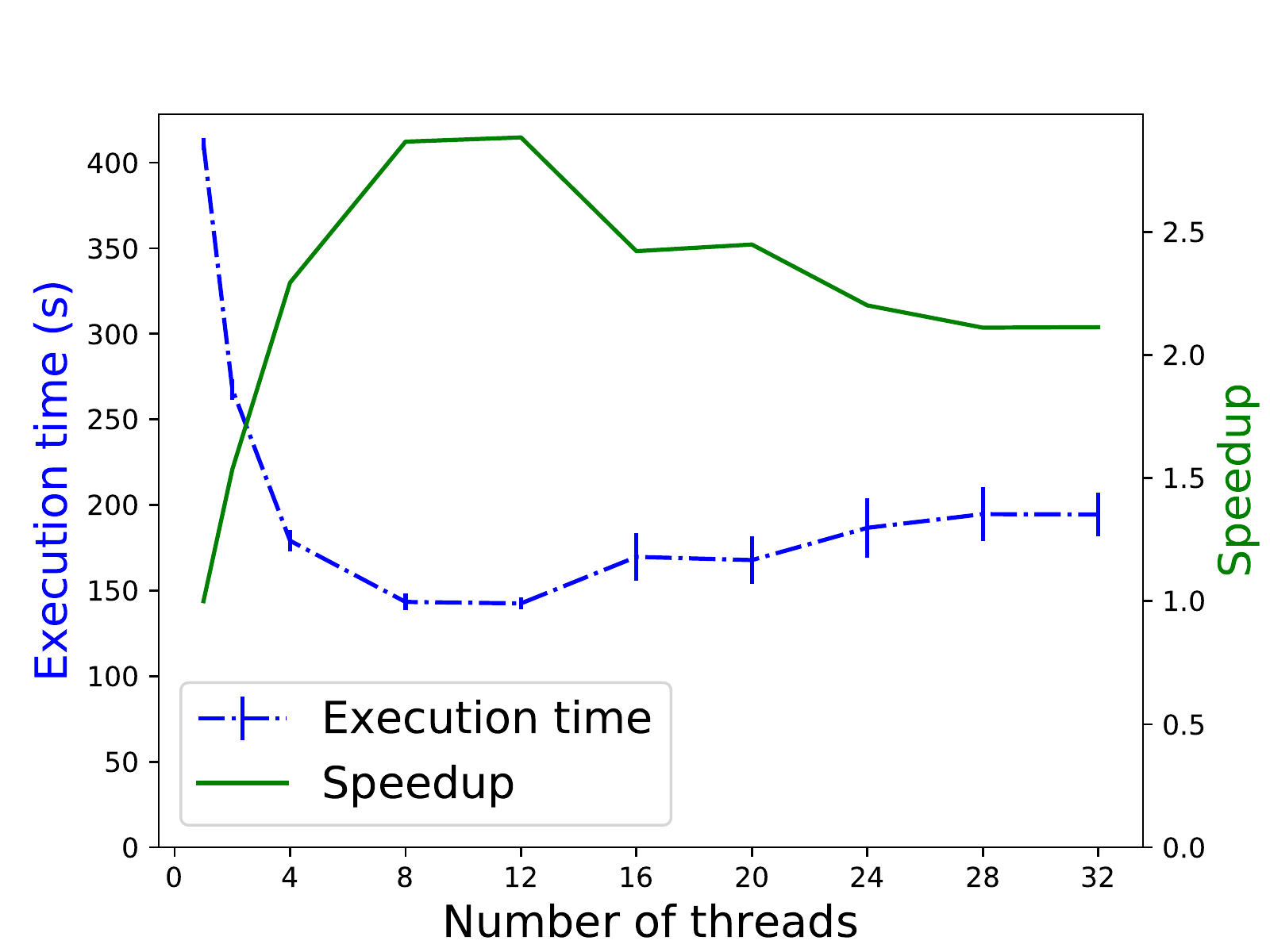}
      \label{fig:xp:9_0_20_16_5}
    }	
		
	\vspace{-1ex}
    \caption{9 constraints, no redundant ones, 16 variables,
      2--36 regions, OpenMP.}
    \label{fig:xp:9_0_20_16}
\end{figure}

\begin{figure}[tp]
    \subfloat[2 dimensions projected, 4 redundant constraints]{
      \includegraphics[width=0.48\textwidth,height=3.8cm]{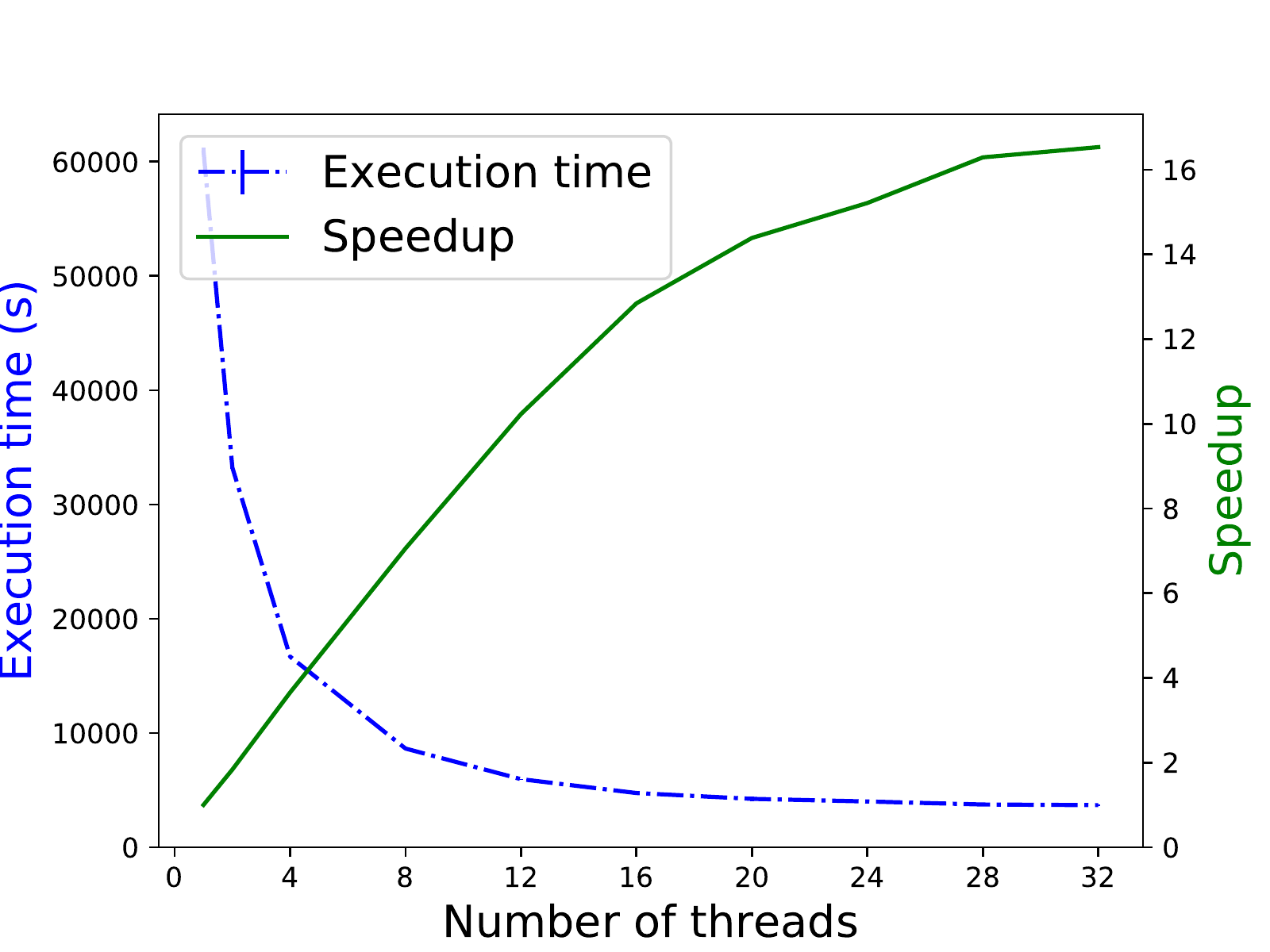}
      \label{fig:xp:24_4_10_6_2}
    }\hfill
    \subfloat[5 dimensions projected, 4 redundant constraints]{
      \includegraphics[width=0.48\textwidth,height=3.8cm]{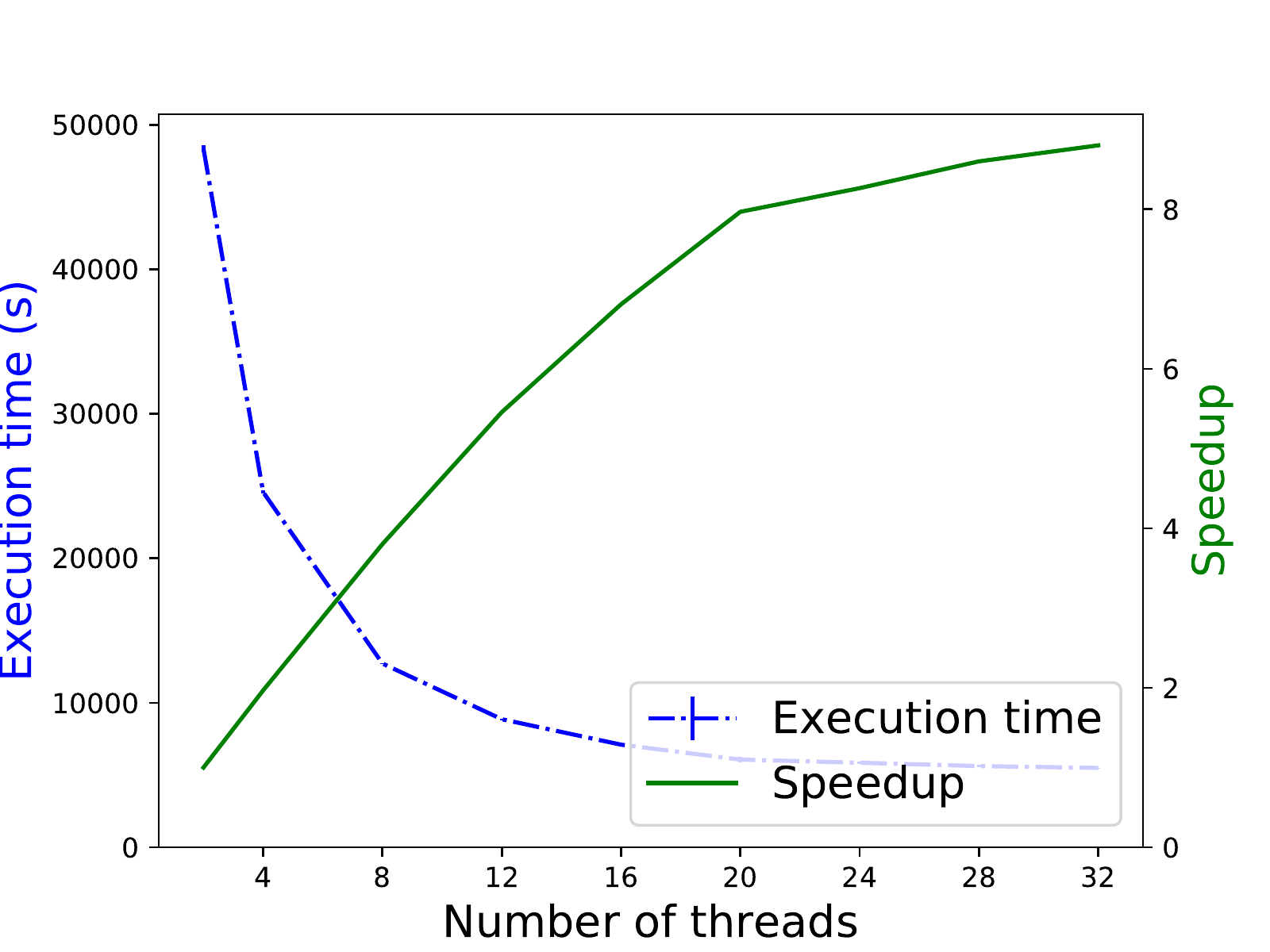}
      \label{fig:xp:24_4_10_6_5}
    }

    \vspace{-1ex}
    
    \caption{24 constraints, 10 variables, 8--764 regions, OpenMP.}
    \label{fig:xp:24_4_10_6}
\end{figure}

\begin{figure}[t]
  \vspace{-1.2em}
  
  \subfloat[OpenMP on Paranoia]{
    \includegraphics[width=.48\textwidth,height=3.8cm]{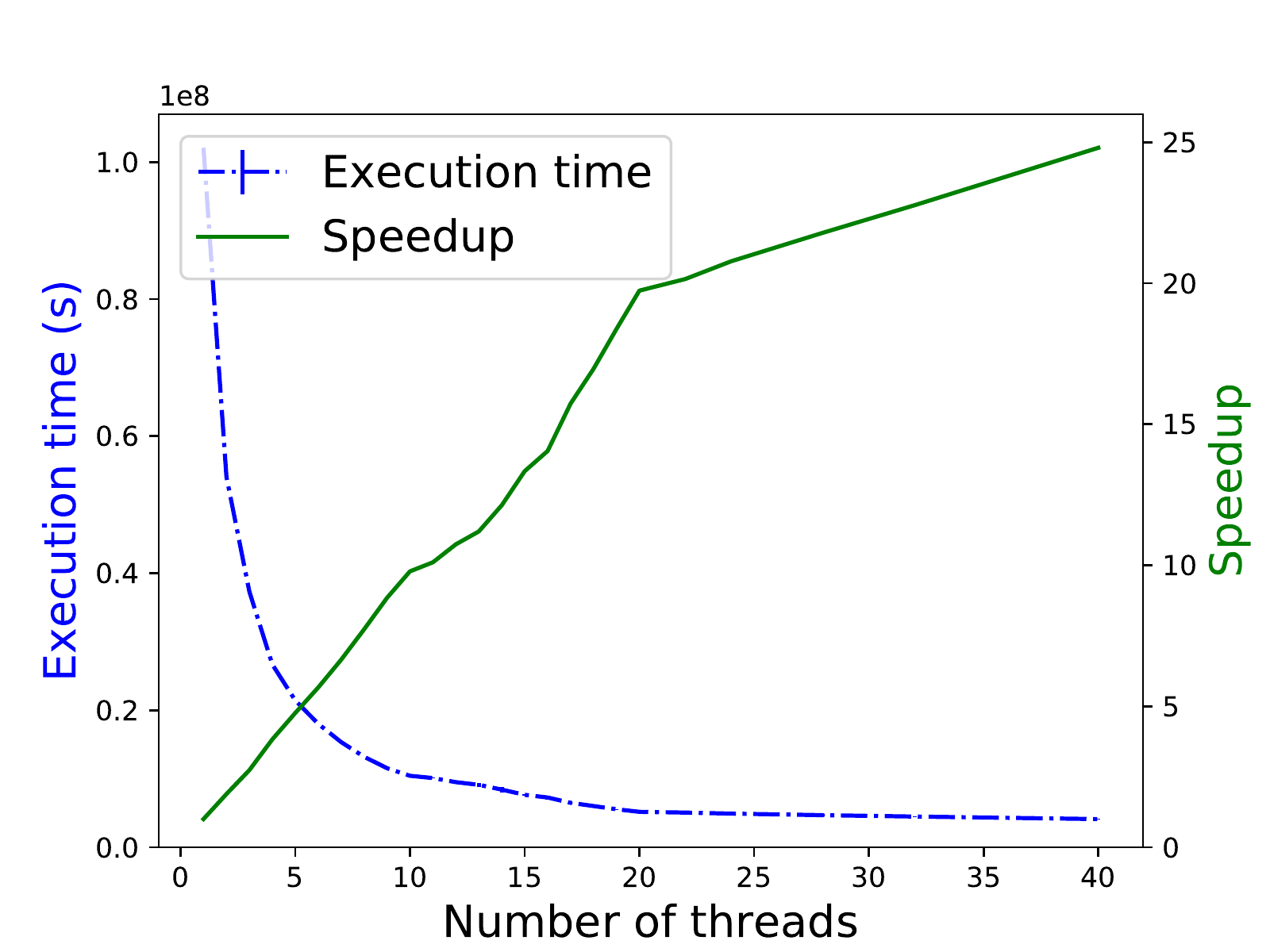}
    \label{fig:xp:120omp}
  }
  \subfloat[TBB on Pressembois]{
    \includegraphics[width=.48\textwidth,height=3.8cm]{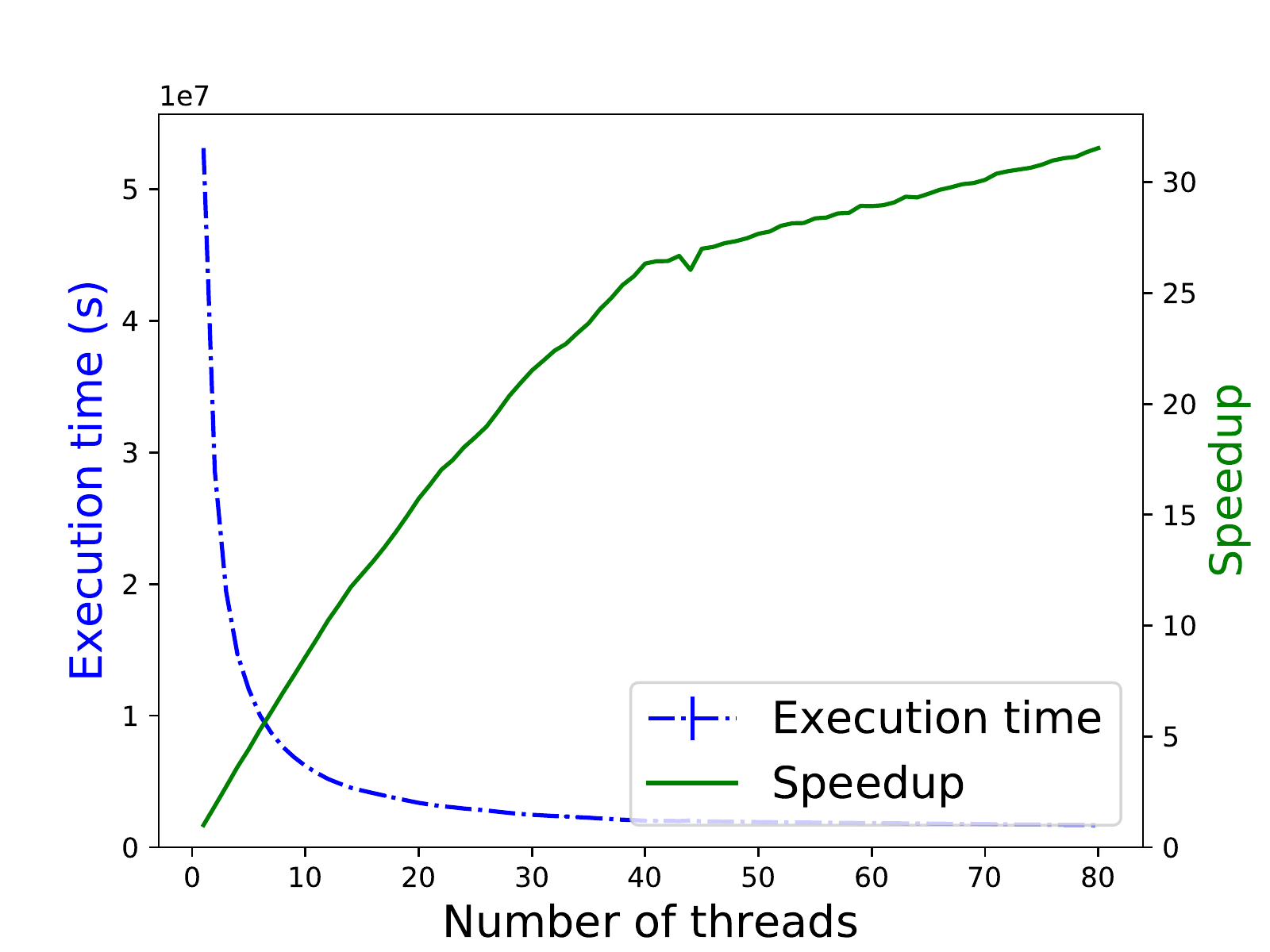}
    \label{fig:xp:120tbb}
  }
  \vspace{-1.2ex}

  \caption{120 constraints, 50 variables, 1 dimension projected, 3459--3718 regions.}
  \label{fig:xp:120}
\end{figure}

\begin{figure}[t]
  \subfloat[1 thread]{
    \parbox{\textwidth}{%
      \includegraphics[width=\textwidth]{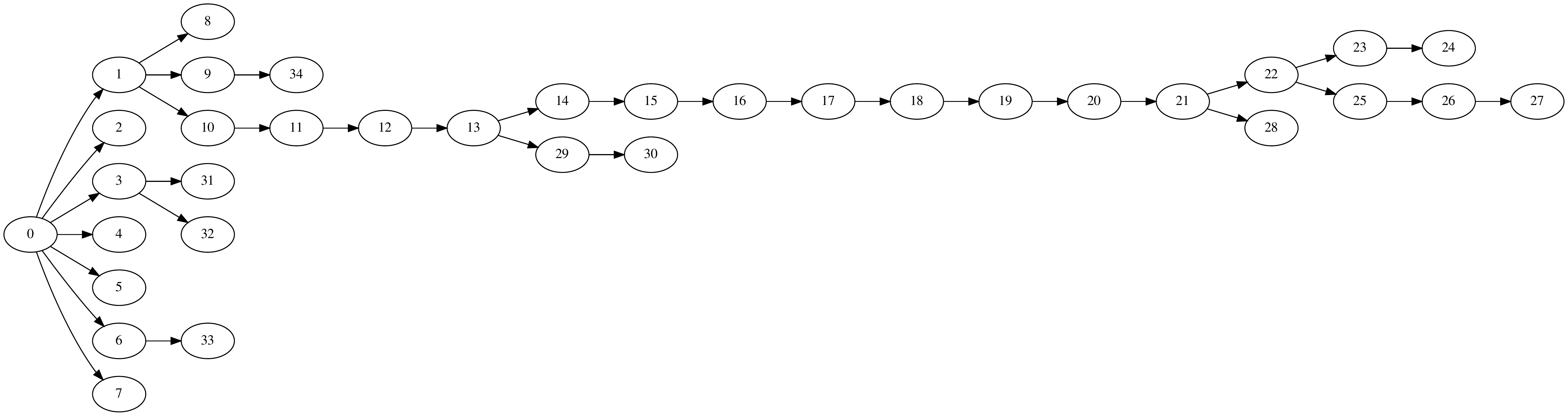}
      \vspace{-2.5em}
    }
  }\\[-1.5em]
  \subfloat[30 threads\label{fig:P_29_15_16_3_0_nproj7_nthreads30}]{
    \includegraphics[width=.59\textwidth]{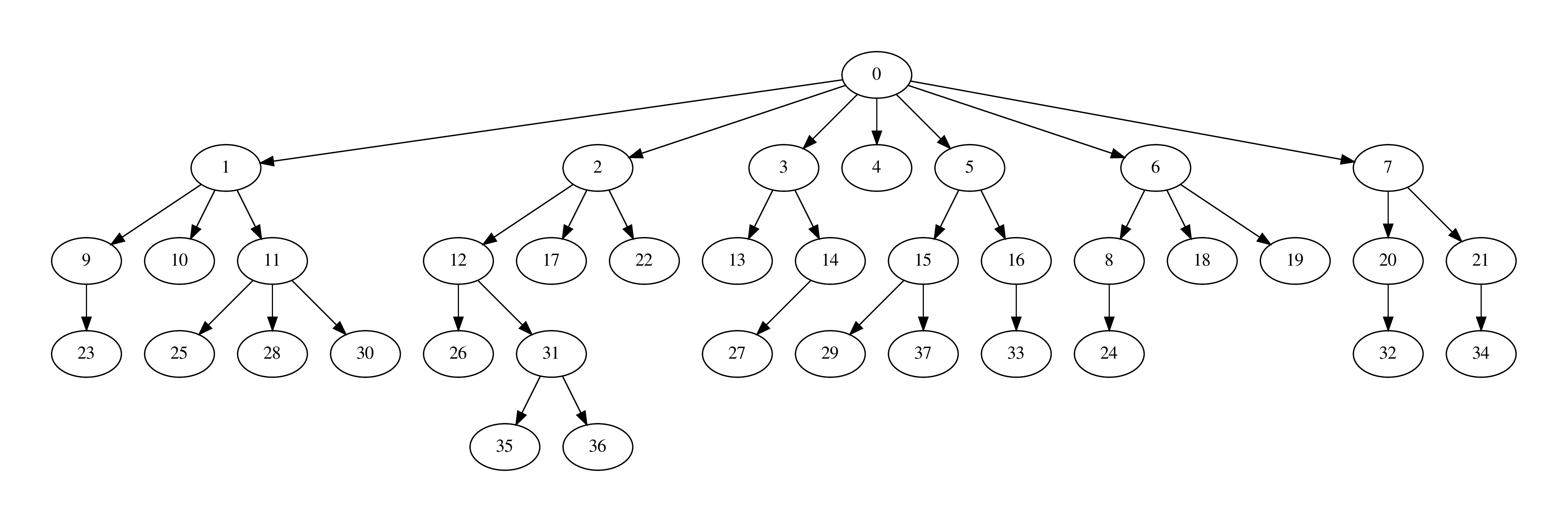}
  }
  \subfloat[Performance]{
        \includegraphics[width=.39\textwidth]{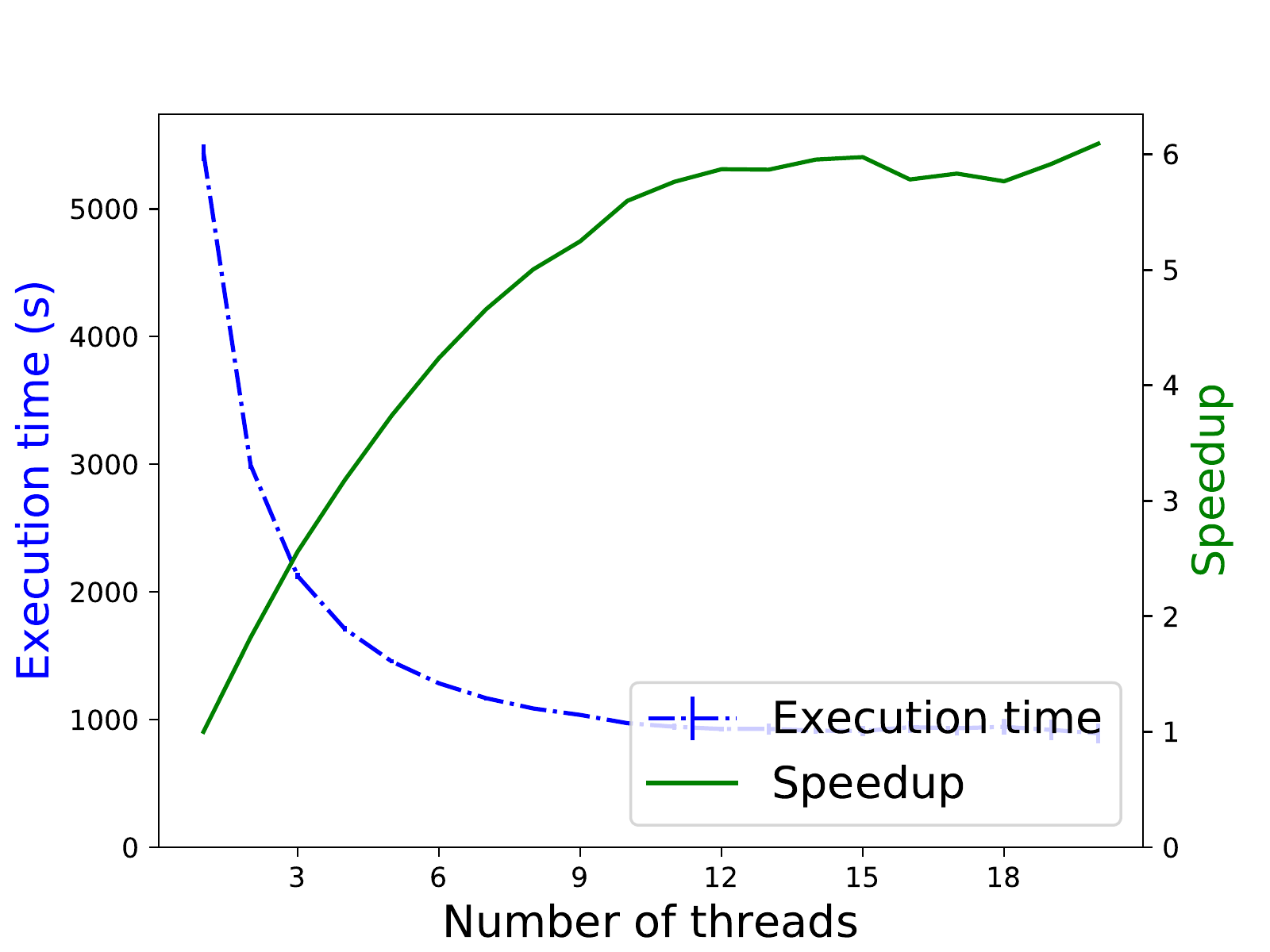}
  }
  \caption{Generation graph of the regions from one polyhedron, computed with 1  and 30 threads.
    The region graphs, depending on overlaps etc., are different; the numbers in both trees have no relationship.}   
  \label{fig:small_polyhedra_spanning_trees}
\end{figure}

We implemented our parallel algorithms in C++, with three alternate schemes selectable at compile-time: no parallelism, OpenMP parallelism or TBB.





All benchmarks were run on the Paranoia cluster of Grid'5000
\cite{grid5000} and on a server called Pressembois.
Paranoia has 8 nodes, each with 2 Intel{\Registered} Xeon{\Registered}
E5-2660v2 CPUs (10 cores, 20 threads/CPU) and 128 GiB of RAM.
Code was compiled using GCC 6.3.1 and OpenMP
4.5 (201511). The nodes run Linux Debian Stretch with a
4.9.0 kernel.
Pressembois has 2 Intel Xeon Gold 6138 CPU
(20 cores/CPU, 40 threads/CPU) and 192 GiB of RAM. It runs a 4.9 Linux
kernel, and we used GCC 6.3.
Every experiment was run 10 times.
The plots presented in this section provide the average and standard deviation.
Paranoia was used for the OpenMP
experiments, whereas Pressembois was used for TBB.

We evaluated our parallel parametric linear programming implementation by using it to project polyhedra, a very fundamental operation.
We used a set of typical polyhedra, with different characteristics:
numbers of dimensions, of dimensions to be projected and of constraints,
sparsity.
Here we present a subset of these benchmarks, each comprising
50 to 100 polyhedra.

On problems that have only few regions, not enough parallelism
can be extracted to exploit all the cores of the machine. For
instance, Figure \ref{fig:xp:9_0_20_16} presents two experiments on 2
to 36 regions using the OpenMP version. It gives an acceptable speed-up on a few cores (up to
10), then the computation does not generate enough tasks to keep the
additional cores busy.
As expected, when the solution has many
regions, computation scales better. Figure \ref{fig:xp:24_4_10_6}
presents the performance obtained on polyhedra made of 24 constraints,
involving 8 to 764 regions, using the OpenMP version. The speed-up
is sublinear, especially beyond 20 cores.

On larger polyhedra, with 120 constraints and 50 variables,
the speedup is close to linear with both OpenMP
and TBB (Fig.~\ref{fig:xp:120}). 
The parallelism extracted from the computation is illustrated by
Fig.~\ref{fig:small_polyhedra_spanning_trees}, on a polyhedron
involving 29 constraints and 16 variables.
Figure~\ref{fig:P_29_15_16_3_0_nproj7_nthreads30} shows the number of
parallel tasks.

\ignore{
The speedup is good for larger problems with many regions,
the number of tasks is larger than the available cores, hence allowing
an efficient parallel computation (Fig.~\ref{fig:large_polyhedra});
it peaks when the problems do not have enough regions  (Fig.~\ref{fig:small_polyhedra});
the limited width of the region graph then limits parallelism
(Fig.~\ref{fig:small_polyhedra_spanning_trees}).
}



\bibliographystyle{plain}
\bibliography{PPLP_ICCS_2019}

\end{document}